\documentclass[iop,apj,tighten]{emulateapj}
\usepackage{apjfonts} 
\usepackage{bm}
\usepackage{amsmath,amssymb}
\usepackage{amstext,mathtools}
\usepackage{here}
\usepackage[breaklinks,colorlinks,citecolor=blue,linkcolor=magenta]{hyperref} 
\usepackage[all]{hypcap} 
\usepackage[usenames,dvipsnames]{color}

\def\be{\begin{equation}}
\def\ee{\end{equation}}
\def\ben{\begin{eqnarray}}
\def\een{\end{eqnarray}}

\DeclareMathAccent{\dot}    {\mathalpha}{operators}{'137} 
\DeclareMathAccent{\ddot}    {\mathalpha}{operators}{'177} 

\shorttitle{On arrival time difference}
\shortauthors{Suyama}

\begin{document}


\title{
On arrival time difference between lensed gravitational waves
and light
}

\author{
Teruaki Suyama\altaffilmark{1}
}

 \affil{$^1$ Department of Physics, Tokyo Institute of Technology, 2-12-1 Ookayama, Meguro-ku, Tokyo 152-8551, Japan}

\begin{abstract}
It is known that geometrical optics no longer applies to the gravitational lensing if
the wavelength of a propagating wave becomes comparable to or larger than the Schwarzshild radius 
of a lensing object.
We investigate the propagation of gravitational waves in wave optics, 
particularly focusing on the difference between their arrival time and the arrival time of light.
We argue that, contrary to the observation in the previous work, gravitational waves never arrive at an observer
earlier than light when both gravitational waves and light are emitted from a same source simultaneously.
\end{abstract}

 \keywords{gravitational waves -gravitational lensing}

 \maketitle

\section{Introduction}
Just as light is bent by gravity, gravitational waves (GWs) are also bent by gravity \citep{MTW}.
This phenomenon, gravitational lensing of GWs, has been acquiring strong 
interest recently \citep{Baker:2016reh, Fan:2016swi, Smith:2017mqu, Jung:2017flg, Oguri:2018muv, Dai:2018enj, Christian:2018vsi, Cremonese:2018cyg,  Liao:2019aqq, Meena:2019ate, Oguri:2019fix, Cusin:2019rmt, Hou:2019dcm, Cremonese:2019tgb, Liao:2020hnx},
especially after the detection of GWs by LIGO \citep{Abbott:2016blz}.
Detection of the lensing of GWs has not been reported yet, but is thought to be a promising discovery in the future
when many merger events occuring at high redshifts are detected,
for instance, by the third generation observatories \citep{Punturo:2010zz, Evans:2016mbw} or pulsar timing array experiments \citep{Sesana:2011zv}. 
Observations of the lensed GWs will provide a completely novel way to probe the compact objects
in the Universe \citep{Takahashi:2003ix}
and matter inhomogeneities on very small scales \citep{Macquart:2004sh, Takahashi:2005ug}.

One prominent feature of the gravitational lensing of GWs in some realistic astrophysical situations
is the wave effect \citep{Schneider, Nakamura:1997sw}.
It is known that when the wavelength becomes comparable to or larger than the Schwarzshild radius of the lens,
geometrical optics breaks down and the wave nature becomes significant \citep{Ohanian:1974ys}.
For instance, with GWs in the LIGO frequency band, 
the wave effect becomes important for the lens mass
\be
M_L \lesssim 300~M_\odot ~{\left( \frac{f}{100~{\rm Hz}} \right)}^{-1},
\ee
where $f$ is the frequency of the GWs.
GWs in the regime given by the above inequality do not follow geodesics and propagate in regions where geodesics do not.
As a result, such GWs provide additional information about the lens, which light does not have \citep{Jow:2020rcy}.

\begin{figure}
\centering
\includegraphics[width=\columnwidth]{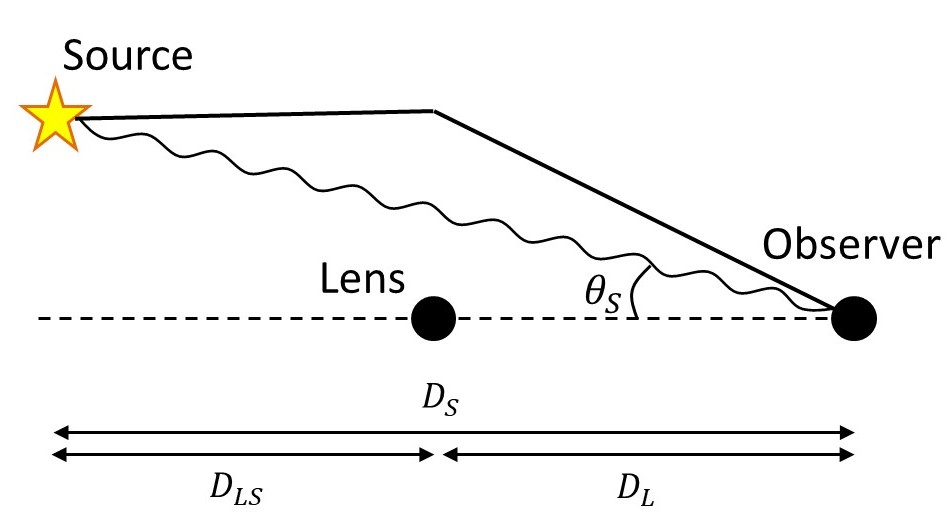}
\label{fig0}
\caption{Schematic picture of the gravitational lensing. Black thick line
is a geodesic, i.e. path of light (secondary path is omitted). Wavy line is a straight path from the source to the observer and is supposed to represent
the propagation of the long-wavelength GWs.}
\end{figure}

Recently, it was claimed in the literature that lensed GWs arrive at the observer earlier than light even if
GWs and light are emitted at the same time from the same source \citep{Takahashi:2016jom}.
An intuitive explanation for this phenomenon is that long-wavelength GWs are less affected by the gravity of the lens
and can propagate straight while the arrival of light is delayed by both geometical deviation from the straight line
and the Shapiro time delay (see Fig.~\ref{fig0}).
However, it is also counterintuitive since the arrival of GWs earlier than light implies that propagation of GWs is superluminal.
In the previous study, it was argued that this issue is not problematic 
in the sense that it is not inconsistent with general relativity.
If this difference of the arrival times is the real effect, it must be taken into account
in the multimessenger observations to correctly interpret the
source properties as well as the lens properties. 
Furthermore, this effect may be also relevant to testing the propagation of the GWs in other theories of gravity. 
The potential importance of this effect in astrophysics and gravitational physics provides a sufficient motivation to reinvestigate this issue.

In this paper, we revisit the propagation of GWs in the framework of the wave optics.
While the difference of the arrival times is defined in terms of the phase of the amplification factor in the previous study, which gives the propagation velocity in the geometrical optics limit, 
we instead consider the front velocity, which gives the correct arrival time for waves consisting of any frequencies.
Our analysis demonstrates that GWs never arrive earlier than light when they are emitted at the same time.
In the first part of the next section, 
we will show that the two events, emission of the GWs at the source
and the reception of the GWs by the observer, are space-like separated
if the effect claimed in the previous study is true.
Then, in the second part of the next section,
we consider the formal expression of the waveform of the lensed GWs and 
give a mathematical proof that the waveform is exactly zero at the
obsever's location before the first light from the source arrives there.
Throughout the paper, the speed of light is set to unity, $c=1$.

\section{Propagation of the lensed GWs}
\subsection{Superluminality of the lensed GWs}
\begin{figure}[t]
\centering
\includegraphics[width=\columnwidth]{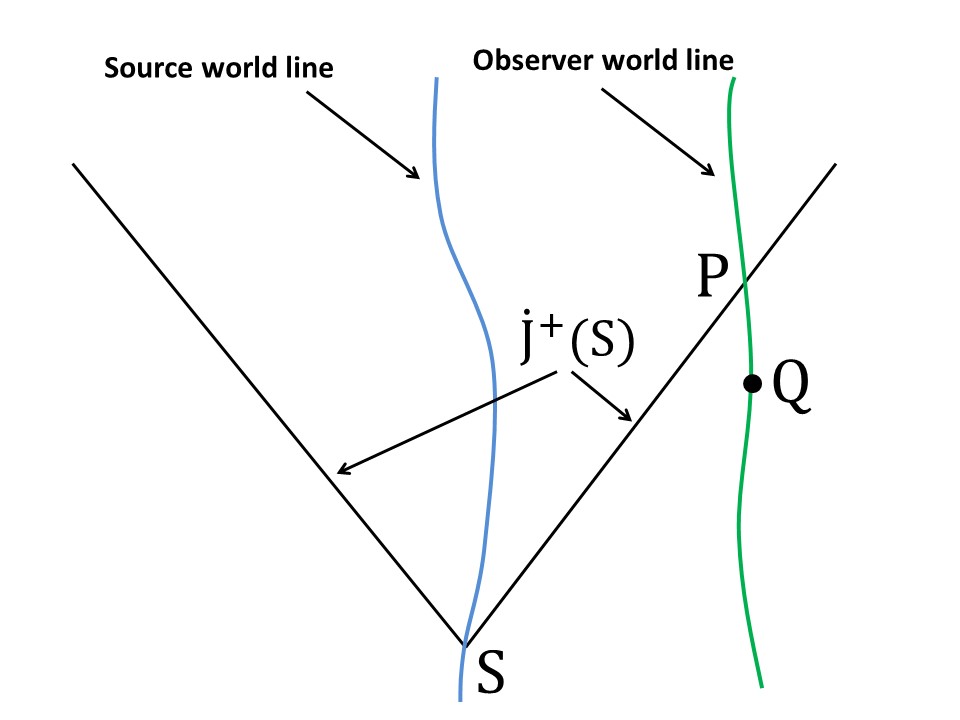}
\label{fig1}
\caption{Schematic picture of the causality in the gravitational lensing. 
Blue curve is a world line of the source and the green curve is a world
line of the observer. Both GWs and light are emitted at a point $S$.
The first light arrives at the observer at a point $P$ which is 
on ${\dot J}^+ (S)$.
}
\end{figure}

We argue in this subsection that earlier arrival of the lensed GWs than light indeed means that propagation of GWs is superluminal.
Let us consider a situation where the point-like source starts emission of both GWs and light (isotropically) at $t=0$.
This event corresponds to a point $S$ in Fig.~\ref{fig1}.
In curved spacetime, it can happen that several lights propagating
different paths arrive at the observer at different times.
Among such lights, we focus on the first light that arrives at the observer.
At point $P$, the world line of the observer intersects the boundary of the causal future of $S$ (${\dot J}^+ (S)$).
It is known that any causal curve connecting $S$ and $P$
is a null geodesic (see for instance \cite{Wald}).
Since the light propagates along a null geodesic, a path of the first light
that arrives at the observer
is on ${\dot J}^+ (S)$.
Now, if the GWs emitted at $S$ arrive at the observer prior to the first light,
the event of the arrival must be outside $J^+(S)$ like a point $Q$ in Fig.~\ref{fig1}.
Thus, the event that the GWs arrive earlier than the first light is not causally connected to $S$.
In this sense, the propagation of the GWs is superluminal.
It is worth mentioning that the discussion in this subsection does not assume Einstein
equations and can be applied to other theories of gravity.
The discussion here suggests that the waveform of the lensed GWs should vanish
outside $J^+ (S)$ when it is computed appropriately,
which is the topic below.

\subsection{Lensed waveform}
In this subsection, by explicitly evaluating the waveform of the lensed GWs, 
we demonstrate that the lensed GWs do not arrive at an observer earlier than light.
This issue is already discussed partially in \cite{Peters:1974gj} in which the propagation of the lensed GWs 
is shown to respect causality in the Born approximation.
Here, we go beyond the previous study by adopting a modern formulation of the waveform 
that is free from the Born approximation and also 
provide a simple argument for the non-superluminal propagation of the GWs. 
To this end, we first give a brief overview of the basic equations of the gravitational lensing 
relevant to our discussion \citep{Nakamura:1999}.
In what follows, 
we ignore the polarization degree of the GWs and the cosmic expansion since
they are not essential to the current purpose.
We denote the GWs by $\phi$.

The presence of the lens object distorts the spacetime from the Minkowski one.
In most astrophysical situations, the distortion is small and it is a good approximation
to write the metric around the lens as
\be
ds^2=-(1+2U) dt^2+(1-2U) d{\vec x}^2.
\ee
Here $U$ is the gravitational potential sourced by the lens object. 
This metric is correct up to first order in $U$. 
GWs and light propagate on this background spacetime.
Wave equation for $\phi$ emitted by the point-like source located at the origin is given by
\be
\bigg[ -(1-4U) \frac{\partial}{\partial t^2}+{\vec \nabla}^2 \bigg] \phi (t,{\vec x})=-4\pi S(t) \delta ({\vec x}),
\ee
where $S(t)$ represents time dependence of the source properties.
If the distance between the source and the lens object is much larger than the wavelength of GWs,
it is known that the solution of the above equation is given by \citep{Schneider, Nakamura:1999}
\be
\phi (t,{\vec x})=\frac{1}{D_S} \int \frac{d\omega }{2\pi} e^{i \omega (D_S-t)} F(\omega, {\vec \theta}_S) {\tilde S}_\omega  \label{sol-phi}
\ee
where ${\tilde S}_\omega$ is Fourier transformation of $S(t)$ and $F(\omega,{\vec \theta}_S)$ is the so-called
amplification factor whose expression under the thin lens approximation is given by
\be
F(\omega,{\vec \theta}_S)=\frac{D_L D_S}{D_{LS}} \frac{\omega}{2\pi i}
\int d^2 \theta \exp \left( i\omega t_d ({\vec \theta},{\vec \theta}_S) \right),
\ee 
where $t_d({\vec \theta},{\vec \theta}_S)$ is the time delay between the lensed light
and the unlensed light\footnote{ 
Validity of the thin lens approximation was investigated in \cite{Suyama:2005mx}.}.
See Fig.~\ref{fig0} for the definition of some symbols.
Explicitly, it is given by
\be
t_d({\vec \theta},{\vec \theta}_S)=\frac{D_L D_S}{2D_{LS}} {|{\vec \theta}-{\vec \theta}_S|}^2-\psi ({\vec \theta}).
\ee 
The first term represents the geometrical time delay, and the second term, which is the lensing potential, 
represents the Shapiro time delay.

Having given the basic equations, let us consider a source which starts emission of GWs at $t=0$.
Namely, $S(t)$ is given by
\be
S(t)=\begin{cases}
s(t) ~~~~~(t \ge 0)\\
0~~~~~~~~~ (t<0).
\end{cases}
\ee
For this source, the GWs given by Eq.~(\ref{sol-phi}) becomes
\begin{align}
\phi (t,{\vec x})=\frac{D_L}{D_{LS}} \int \frac{d\omega }{2\pi} e^{i \omega (D_S-t)}
\frac{\omega}{2\pi i} \int d^2 \theta \exp \left( i\omega t_d ({\vec \theta},{\vec \theta}_S) \right) \nonumber \\ 
\times \int_0^\infty dt' e^{i\omega t'} s(t').
\end{align}
Integration over $\omega$ yields
\be
\phi (t,{\vec x})=\frac{D_L}{D_{LS}} \frac{\partial}{\partial t}
\int d^2 \theta \int_0^\infty dt'~s(t') \delta (D_S-t+t'+t_d ({\vec \theta},{\vec \theta}_S)),
\ee
where $\delta (x)$ is Dirac's delta function.
Since the range of integration of $t'$ is $t' \ge 0$,
the integrand becomes non-vanishing only for $t \ge t_{\rm min}$,
where $t_{\rm min}$ is given by
\be
t_{\rm min}=D_S+\underset{\vec \theta}{\rm min} \{ t_d ({\vec \theta},{\vec \theta}_S) \}. \label{t-min}
\ee
Since ${\vec \theta}$ that minimizes $t_d ({\vec \theta},{\vec \theta}_S)$ is 
a solution of the lens equation \citep{Schneider},
$t_{\rm min}$ is nothing but the first time when the light emitted from the source at $t=0$ arrives at the observer.
Thus, we have shown that GWs never arrive earlier than light when two are emitted
at the same time at the same place.
This conclusion is opposite to the observation made in \cite{Takahashi:2016jom}.
The source of this discrepancy may be ascribed to the fact that the time delay in the previous study
is defined in terms of the phase of the GWs in the frequency domain, 
which does not necessarily coincide with the front velocity.
Typical time delay caused by the gravitational lensing is the order of the Schwarzchild radius
of the lens object.
Thus, the notion of the arrival time for waves whose wavelength is larger than the Schwarzchild radius
becomes ambiguous when the time delay of interest is the order of the Schwarzshild radius.
On the other hand, things become much clearer if one studies the waveform in the time domain,
as it has been done here. 

In the previous studies \citep{Takahashi:2016jom, Cremonese:2018cyg}, it was argued that observational determination of the arrival time difference in terms of the phase of the amplification factor are potentially useful in multimessenger astronomy and tests of GW propagation.
In \cite{Cremonese:2019tgb}, a new idea was proposed that the arrival time difference can be used to determine the Hubble constant within the accuracy better than the existing measurements.  
The analysis in this section suggests that it is not appropriate to use the phase velocity and it is worth revisiting these studies to clarify how the observational consequences are changed accordingly. 

\begin{figure}[t]
\centering
\includegraphics[width=\columnwidth]{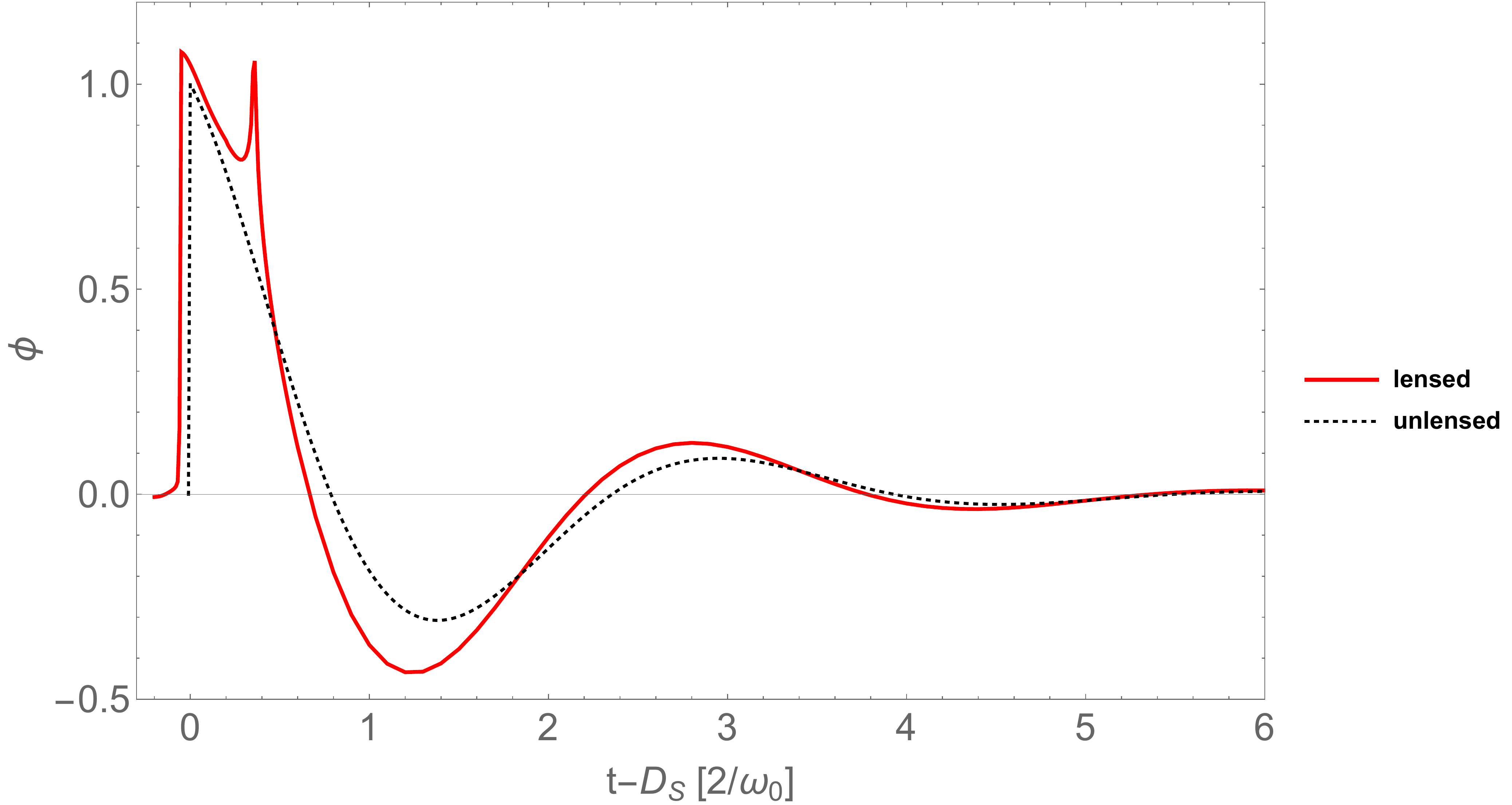}
\label{fig2}
\caption{Waveform lensed by a point mass in the time domain (red thick curve). 
The values of the parameters are $(\Gamma,2GM_L) =(0.4 \omega_0, 0.2 \omega_0^{-1})$.
As a reference, unlensed waveform is also presented as a black dotted curve.
}
\end{figure}

\section{An example of the lensed waveform in time domain}
In this section, we compute a waveform lensed by a point mass for a simple source 
as an illustration of how a waveform of the GWs is deformed
by the gravitational lensing in the wave optics.
We consider the following form of the source $S(t)$
\be
S(t)= \theta (t) e^{-\Gamma t} \cos (\omega_0 t),
\ee
where $\theta (t)$ is the Heaviside function and $\Gamma, \omega_0$ are positive constants.
The Fourier transform of this function is given by
\be
{\tilde S}_\omega=\frac{\Gamma -i\omega}{{(\Gamma-i\omega)}^2+\omega_0^2}. \label{point-S}
\ee
Analytic form of the amplification factor for the point mass lens with its mass $M_L$ exists and it is given by
\citep{Schneider}
\be
F(\omega,{\vec \theta}_S)=e^{\frac{\pi}{4}w+\frac{iw}{2} \ln \left( \frac{w}{2} \right)}
\Gamma \left( 1-\frac{iw}{2} \right) 
{_1F_1}\left( \frac{iw}{2},1;\frac{i}{2}wy^2 \right), \label{point-amp}
\ee
where $w \equiv 4GM_L \omega$ and $y\equiv \sqrt{\frac{D_S D_L}{4GM_L D_{LS}}}\theta_S$.
The lensed waveform can be computed by plugging 
Equations (\ref{point-S}) and (\ref{point-amp})
into Equation (\ref{sol-phi}).

Fig.~\ref{fig2} shows both lensed (red thick curve) and unlensed (black dotted curve)
waveforms in the time domain.
The values of the parameters are $(\Gamma,2GM_L) =(0.4 \omega_0, 0.2 \omega_0^{-1})$.
The lensed waveform exhibits a few features that may warrant mentioning.
First, it is verified that the wave vanishes until the time $t_{\rm min}$ given by Eq.~(\ref{t-min}), 
which corresponds to the first arrival time of light, 
and the wave sharply rises at $t_{\rm min}$ as it has been shown in the previous section.
A tiny tail seen prior to the initial rise is a numerical artifact.
Second, magnification due to lensing makes the height of the first peak larger than unity,
though not by a significant amount for the parameter values in the present case.
Third, there is a second peak. 
This is mainly caused by GWs which propagate along a secondary path of the lens equation.

\section{Summary}
It has been proposed in the previous study that the lensed GWs arrive at an observer 
earlier than light even when both GWs and light are emitted simultaneously.
Because of the potential impacts of this claim on both astrophysics and gravitational physics,
it is important to reconsider this observation.
We argued that the claimed effect means superluminal propagation of the GWs in the 
sense that two events, emission of the GWs by the source and reception of the GWs by the observer,
are space-like separated.
We then showed, by explicitly evaluating the waveform of the lensed GWs, that 
GWs never arrive earlier than light.
This conclusion holds independently of the density profile of the lens object as well as the waveform of the GWs.
Our finding may be used to constrain the emission time difference of the GWs and light
from the same source when the lensed GWs and light are detected.

\acknowledgments{
TS thanks to Saul Herwitz and Ryuichi Takahashi for useful comments.
TS is supported by JSPS Grant-in-Aid for Young Scientists (B) No.15K17632,  
by the MEXT Grant-in-Aid for Scientific Research on In-novative Areas No.15H05888, No.17H06359, No.18H04338, and No.19K03864.
}

\bibliographystyle{yahapj}
\bibliography{draft}

\end{document}